\documentstyle[epsfig]{aipproc}

\begin{document}
\title{
The Early Formation, Evolution and \\ Age of the Neutron-Capture
Elements in the Early Galaxy\thanks{To appear 
in the {\it Proceedings of the 20th Texas Symposium on
Relativistic Astrophysics}, J. C. Wheeler \& H. Martel (eds.)}
}
\author{John J. Cowan$^*$, Christopher Sneden$^\dagger$ 
and James W. Truran$^{\P}$}
\address{$^*$Department of Physics and Astronomy, University of Oklahoma,
Norman, OK 73019\\
$^{\dagger}$Department of Astronomy, University of Texas,
Austin, TX 78712\\
$^\P$Department of Astronomy and Astrophysics, University of Chicago,
Chicago, IL 60637}

\maketitle

\begin{abstract}
Abundance observations  indicate the presence of
rapid-neutron capture (i.e., {\em r}-process) elements in old Galactic halo
and globular cluster stars. These observations demonstrate that
the earliest generations of stars in the Galaxy, responsible
for neutron-capture synthesis and the progenitors of the halo stars,
were rapidly evolving.
Abundance comparisons among several halo stars show that the heaviest
neutron-capture elements (including Ba and heavier) are
consistent with  a scaled solar system {\em r}-process abundance distribution,
while the lighter such elements do not conform to the solar pattern.
These comparisons suggest two  {\em r}-process sites or at least two different
sets of astrophysical conditions.
The large star-to-star scatter observed in the neutron-capture/iron ratios at
low metallicities -- which disappears with increasing [Fe/H] --
suggests an early, chemically unmixed and inhomogeneous Galaxy.
The  stellar abundances indicate a change from the {\em r}-process to the slow
neutron capture (i.e., s-) process at  higher metallicities in the Galaxy.
The detection of thorium in halo and globular cluster stars
offers a promising, independent age-dating technique that can put lower limits
on the age of the Galaxy.
\end{abstract}

\section*{INTRODUCTION}

In this paper we briefly review some of the important abundance 
trends for the slow- or rapid-neutron 
capture elements. 
We focus on how these neutron-capture elements can be employed to 
(1) study the nature of the progenitors and the nucleosynthesis history
in the early Galaxy, (2) explore   the 
chemical evolution of the Galaxy, and 
(3) obtain radioactive age estimates for the oldest stars,
which in turn puts limits on the age of the Galaxy and the universe.

\section*{Neutron-Capture Abundances}
 
\begin{figure}[b!] 
\centerline{\epsfig{file=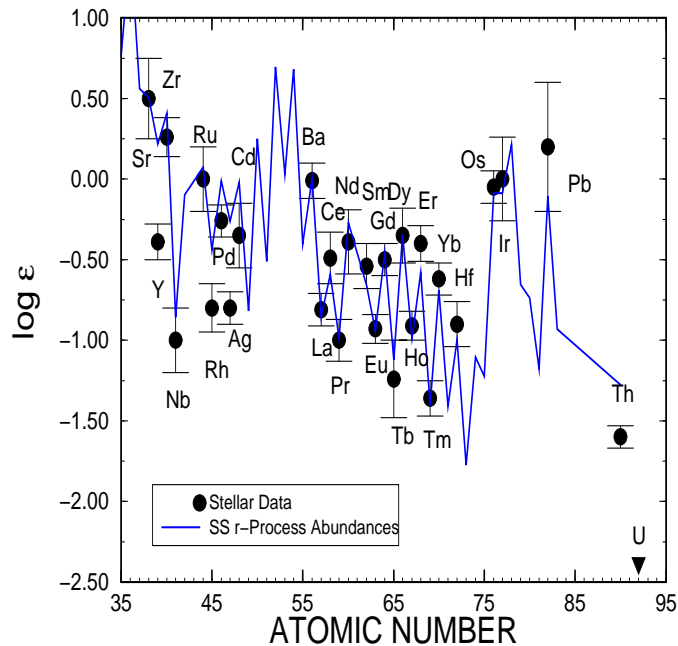,height=3.5in,width=3.5in}}
\vspace{10pt}
\caption{Neutron-capture abundances in CS 22892--052 compared with 
a scaled solar system {\em r}-process distribution (solid line).}
\label{fig1}
\end{figure}

Extensive abundance studies have been made of the ultra-metal-poor ($\equiv$ UMP,
[Fe/H] = --3.1) but neutron-capture rich halo 
star CS~22892--052 \cite{cow95,sne96,sne00}.
In Figure~\ref{fig1} we show the {\em n}-capture abundances as 
determined by Sneden et al.\cite{sne00} compared to a scaled solar system
{\em r}-process abundance distribution. As has been noted previously, the
upper end of the stellar abundance distribution ({\it i.e.}, Ba and above)
is in very close agreement with the solar system {\it r}-process curve. It has been 
seen only recently, and so far only in this star, that the lighter 
{\em n}-capture abundances between Zr and Ba ({\it e.g.}, Nb, Pd, and Ag) 
do not lie on the same solar curve.
This lends support to previous suggestions that there may be two 
sites for the {\em r}-process, one for the heavier and one for the lighter
{\em n}-capture elements \cite{was96}. 
It is unclear whether both of those sites might be supernovae occurring 
at different frequencies \cite{was00} or neutron star binaries 
\cite{ros99} or some combination of those. 
Alternatively, it has been proposed that both 
ends of the abundance distribution could be synthesized in 
different regions of the same  neutron-rich jet of a core-collapse
supernova \cite{cam01}.
 
\section*{Abundance Trends in the Galaxy}

Observations of {\em n}-capture abundances in 
a wide range of Galactic, including metal-poor halo and
disk,  stars have now been made over a range of 
metallicities. These data demonstrate several interesting 
abundance  trends in the Galaxy.

\subsection*{Scatter in the Early Galaxy}

\begin{figure}[b!] 
\centerline{\epsfig{file=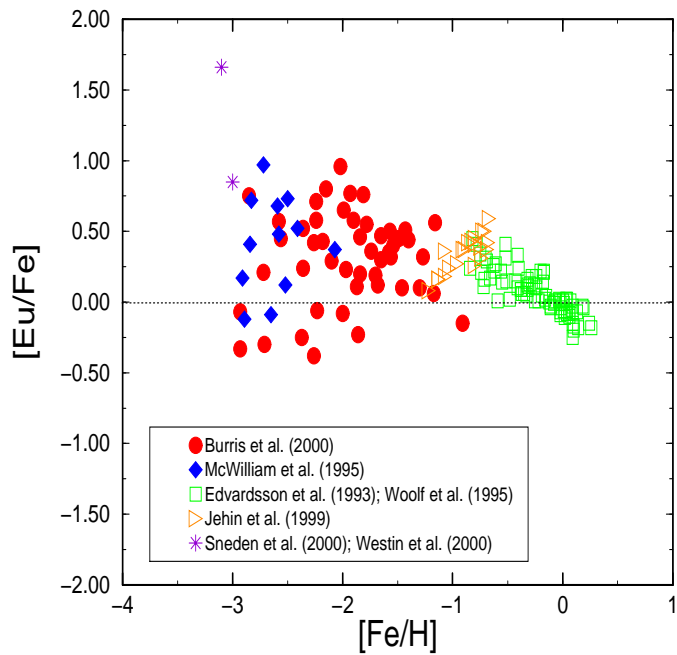,height=3.5in,width=3.5in}}
\vspace{10pt}
\caption{[Eu/Fe] vs. metallicity for Galactic halo and disk stars. 
The dotted line indicates the solar value.}
\label{fig2}
\end{figure}

Earlier work by Gilroy {\it et al.}\cite{gil88}
first demonstrated that the stellar abundances of 
{\em r}-process elements with respect to iron, particularly Eu/Fe, showed 
a large scatter at low metallicities. This scatter appeared to diminish
with increasing metallicity. A more extensive study by 
Burris {\it et al.}\cite{bur00}
confirmed the very large star-to-star  scatter in the early Galaxy, 
while studies of 
stars with higher metallicities -- mostly disk stars -- 
\cite{edv93,wol95} show little scatter. 
In Figure \ref{fig2} we plot the data from a number of surveys 
\cite {bur00,edv93,wol95,mcw95,jeh99},  along with detailed
abundance determinations from several single stars \cite{sne00,wes00}. 
These studies,
which cover a metallicity range 
--3.5$\le$[Fe/H]$\le$+0.5
and include large numbers of stars,  had  
attempted to minimize observational errors. 
The star-to-star 
scatter illustrated in this figure 
can be explained as the result of individual nucleosynthetic
events ({\it e.g.}, supernovae) \cite{gil88,bur00} and strongly suggests  
an early, unmixed, chemically inhomogeneous Galaxy.
(See also \cite{sne01} for further discussion.) It should be noted
that while the absolute levels of {\em n}-capture/Fe abundances vary widely,
the relative abundances are similar in all of the very metal-poor halo
stars.

One other important trend is notable in Figure \ref{fig2}.
At higher metallicities, particularly for [Fe/H] $\simeq$ --1,    
the values of [Eu/Fe] tend downward. 
This demonstrates  clearly the effect of increasing iron production,
presumably from Type Ia supernovae, at higher Galactic metallicities
\cite{bur00}. At very low metallicities high mass (and 
rapidly evolving) Type II supernovae contribute to Galactic iron production.
The onset of the bulk of iron production from Type Ia supernovae
(with longer evolutionary timescales due to lower
mass progenitors) occurs only at higher [Fe/H]  and later Galactic times.

\subsection*{Chemical Evolution of the {\em r}- and {\em s}-Process}

\begin{figure}[b!] 
\centerline{\epsfig{file=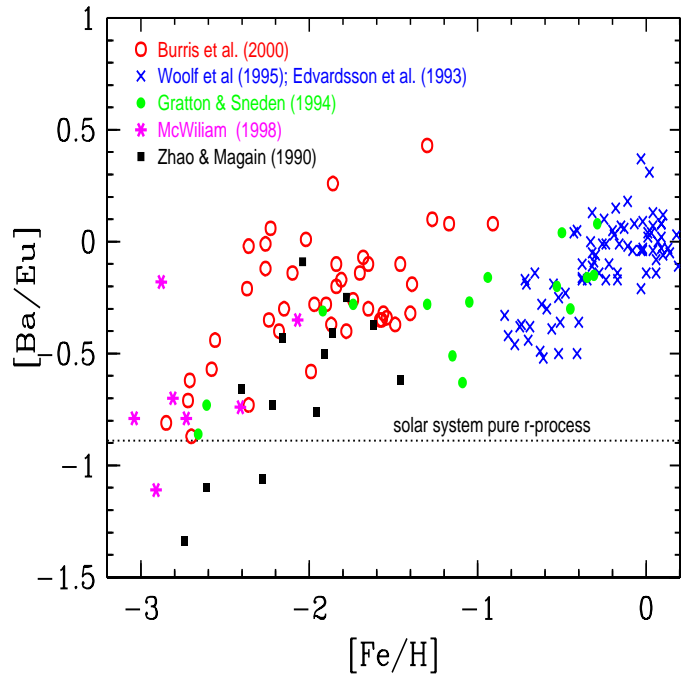,height=3.5in,width=3.5in}}
\vspace{10pt}
\caption{[Ba/Eu] vs. metallicity for Galactic halo and disk stars.
}
\label{fig3}
\end{figure}

The abundances observed for the elements in CS 22892--052 and other 
UMP  halo stars demonstrate the early onset of the {\em r}-process
in the Galaxy. These results (see Figure 1) also 
confirm earlier predictions \cite{tru81} that elements 
synthesized by the {\em s}-process in the solar system ({\it e.g.}, Ba) 
were formed solely in the {\em r}-process early in the history of the Galaxy.
Further confirmation of   
this  early Galactic dominance of the {\em r}-process 
is seen in Figure \ref{fig3},  
where we plot [Ba/Eu] as a function of [Fe/H]. 
We have utilized a combination  of data sets\cite{bur00,edv93,wol95,gra94,mcw98,zha90},  
including some 
shown in Figure 2,  to produce this new figure. 
It is clearly seen in Figure 3  that at the lowest metallicities the 
stellar Ba/Eu ratios cluster around the solar system (pure) {\em r}-process 
value.
Eu is almost exclusively an {\em r}-process element, but
Ba is produced predominantly in solar system material by the
{\em s}-process in low mass (1-3 M$_{\odot}$) AGB stars\cite{bur00}.
At the lowest metallicities early in the history of the Galaxy, the halo
stars show the products only of {\em r}-process nucleosynthesis 
from rapidly evolving (with short stellar evolutionary timescale) 
progenitors  typical of,  
for example,  Type~II supernovae. 
As the metallicity grows larger, the ratio of Ba/Eu rises  due
to the increased production of Ba (in the {\em s}-process) but not the 
{\em r}-process
element Eu.
The transition between
a pure {\em r}-process production of Ba and production dominated
by the main {\em s}-process occurs between --3 $<$ [Fe/H] $<$ --2, 
with most stars consistent with  the (total) solar value of [Ba/Eu] for 
metallicities [Fe/H] $\geq$ --2. 
The delay in the onset of the {\em s}-process with respect to the 
{\em r}-process is consistent with the  longer stellar
evolutionary timescales typical of low-mass (1--3 M$_{\odot}$) stars
thought to be the site for {\em s}-process synthesis.
It is interesting to note that the onset of the bulk of the main 
{\em s}-process 
in the Galaxy at  
[Fe/H] $\simeq$ --2 occurs at a lower metallicity, and likely  an 
earlier time, than the bulk of the iron production from Type Ia 
supernovae at [Fe/H] $\simeq$ --1, as discussed above.
We note further that 
while Ba has been commonly used for these abundance studies, 
in the future La may prove to be a 
more reliable indicator of the Galactic evolution of the {\em s}-process
\cite{sne01}.

\section*{Chronometric Ages}

The detection of the radioactive element thorium in halo
stars such as CS~22892--052 (see Figure 1) has provided the exciting
opportunity of directly determining stellar ages. This technique relies
upon comparing the observed stellar abundances with estimates of 
the initial abundance of the radioactive element. To minimize systematic
errors, ratios of Th to Eu (produced almost exclusively in the 
{\em r}-process) are usually employed for these age determinations. 
Cowan {\it et al.}\cite{cow99} and Westin {\it et al.}\cite{wes00} 
obtained an average (minimum) age of 13.8 Gyr for  
the UMP stars CS~22892--052 and HD~115444 
by comparing their observed Th/Eu abundances with the solar system
ratio. This value represents a lower limit on their ages since Th has partially
decayed over the last 4.5 Gyr. Improvements in these age estimates were 
then made by determining the initial (zero-decay) values of Th/Eu in
the same calculations that reproduced the observed stable {\em n}-capture 
abundance distributions for both of these stars. 
The stable lead and bismuth solar system isotopic abundances were then
employed to determine the most reliable  mass formulae for 
predicting the properties of nuclei far from stability, 
critical for the theoretical {\em r}-process abundance calculations. 
Utilizing these constraints to obtain zero-decay Th/Eu abundances, 
an average age for CS~22892--052 and
HD 115444 of 15.6 Gyr, with an estimated uncertainty of $\simeq$ 4 Gyr,
was obtained \cite{cow99,wes00}.

In addition to studies of halo stars there have been recent observations of
the globular cluster M 15 \cite{sne0b}. Detailed abundance determinations
confirm that the heavier {\em n}-capture abundances are consistent with the
scaled solar system {\em r}-process curve. Further,  the detection of thorium
in several of the globular stars has allowed a  chronometric age 
estimate of 14 Gyr to be determined for this system \cite{sne0b}. 

Uranium, another long-lived radioactive element, can also serve as a 
chronometer. This element has not been detected to date in CS~22892--052 or 
in HD~115444, but has been found for the first time in the star
CS~31082--001 \cite{cay01}. Combining both chronometers Th and U,
in conjunction with several stable heavy elements, 
an age of 12.5 $\pm$ 3 Gyr  has been estimated  for this 
star. While this technique still has some  
observational and theoretical uncertainties associated with it 
\cite{cow01},  it offers great promise.
In particular 
such chronometric  age estimates  of  UMP  stars
are independent of, and consequently  avoid the large uncertainties in,
Galactic chemical evolution models.
Finally, we note that the radioactive age determinations of these oldest
halo stars put meaningful constraints on both Galactic and 
cosmological age estimates.

\section*{ACKNOWLEDGMENTS}

We thank all of our colleagues who have collaborated with us on various
studies of $n$-capture elements in halo stars.
This research has received support from NSF grants
AST-9986974 to J.J.C.,
AST-9987162 to C.S.
and
from DOE contract B341495 to J.W.T., and from the Space
Telescope Science Institute grant GO-8342.

\end{document}